\documentclass[review, a4paper, 3p,12pt]{elsarticle}
\pdfoutput=1 

\usepackage{upgreek}
\usepackage{amsmath}
\journal{Nuclear Inst. and Methods in Physics Research, A}

\begin{document}
\begin{frontmatter}

\title{Liquid-organic time projection chamber for detecting low energy antineutrinos}

\author[1,2]{T.~Radermacher\corref{cor1}}
\ead{radermacher@physik.rwth-aachen.de}
\author[1,2]{J.~Bosse}
\author[1]{S.~Friedrich}
\author[1,2]{M.~Göttsche}
\author[1]{S.~Roth}
\author[1,2]{G.~Schwefer}
\cortext[cor1]{Corresponding author}
\affiliation[1]{organization={III. Physikalisches Institut B, RWTH Aachen University},
addressline={Sommerfeldstraße 16}, 
city={Aachen},
postcode={52074},
country={Germany}}
\affiliation[2]{organization={Aachen Institute for Advanced Study in Computational Engineering Science (AICES), RWTH Aachen University},
addressline={Schinkelstraße 2},
city={Aachen},
postcode={52056},
country={Germany}}

\begin{abstract}
The MeV region of antineutrino energy is of special interest for physics research and for monitoring nuclear nonproliferation. Whereas liquid scintillation detectors are typically used to detect the Inverse Beta Decay (IBD), it has recently been proposed to detect it with a liquid-organic Time Projection Chamber, which could allow a full reconstruction of the particle tracks of the IBD final state. We present the first simulation-based statistical analysis of the expected signatures. Their unequivocal signature could enable a background-minimized detection of electron antineutrinos using information on energy, location and direction of all final state particles. We show that the positron track reflects the antineutrino's vertex. It can also be used to determine the initial neutrino energy. In addition, we investigate the possibility to reconstruct the antineutrino direction on an event-by-event basis by the energy deposition of the neutron-induced proton recoils. Our simulations indicate that this could be a promising approach which should be further studied through experiments with a detector prototype.
\end{abstract}


\end{frontmatter}

\section{Introduction}
Thanks to important advances in detector technology and analysis techniques, experimental neutrino research has become a field of precision physics that continues to push boundaries of sensitivity and accessible energy.
Over the last years, the $\mathrm{MeV}$ region of neutrino energy has been of particular interest, e.g. for the investigation of reactor, solar and geoneutrinos.

In addition, the energy domain below $10\,\mathrm{MeV}$ can be useful for nuclear monitoring purposes where antineutrinos are produced by $\upbeta$-decaying fission products. With regard to reactors, a number of studies, for instance \cite{Carr2019, Christensen2014},  have examined antineutrino monitoring to observe whether reactors are operational, or even to deduce parameters such as fuel burnup or power. In such applications, detectors would typically be placed in close vicinity to the reactor building. Another project \cite{Watchman2015} looks at the potential for antineutrino monitoring to detect undeclared reactors from some distance, a significantly more challenging task.

For nuclear waste, both in interim storage facilities and in future geological repositories, verification activities are required to control the non-diversion of spent fuel which contains plutonium. For the case that other verification techniques fail, antineutrino monitoring could act as a redundancy without common failure modes \cite{Brdar2017,Wittel2020}. The main challenge will be a low neutrino flux intensity.

Antineutrinos in the $\mathrm{MeV}$-energy range can be detected via their interaction with a proton producing a neutron and a positron, which had also been used for the discovery of the neutrino.
This reaction, called Inverse Beta Decay (IBD), has the advantage of a clear final state signature consisting of a prompt signal of the two $511\,\mathrm{keV}$ photons from the positron annihilation and a delayed signal from the absorption of the neutron in a nucleus. 
Selecting this event signature reduces background sources to a large extent.

Typically, organic scintillation detectors are used to measure the energy depositions of the prompt and delayed signals. For high fluxes, e.g. adjacent to a nuclear reactor, no extensive shielding is necessary \cite{Prospect2018, Stereo2020}. 
It has been shown that using IBD is also effective enough for measuring moderate neutrino fluxes in some distance to the reactors in neutrino oscillation experiments \cite{Kerret2022,An2016,Ahn2010}, though requiring much larger detectors and a more significant overburden that shields against cosmic radiation. 
Lastly, the same method even allows the measurement of the very small flux of geoneutrinos that has been successfully performed by Borexino \cite{BorexinoGeoneutrino2020}.
Here, not only very large amounts of overburden are necessary, but also further measures against background sources such as veto counters and shielding against intrinsic radiation of the surrounding rocks are required. In addition, the scintillator needs to be of high purity to avoid radioactive decays inside of the detector.

These elaborate precautions against background are needed due to the fact that in liquid scintillation detectors the final state of the IBD reaction is not reconstructed in detail. 
Instead, the measurements are based on the three following observables of the event: The prompt energy from the positron annihilation, the delayed energy caused by the neutron absorption and the time between both energy depositions. For instance, IBD events cannot be distinguished from events involving electrons and neutrons (e.g. from cosmogenically produced ${}^{9}\mathrm{Li}$ or ${}^{8}\mathrm{He}$) as annihilation photons cannot be identified separately, because the prompt energy of the event, the kinetic energy of positron and the energy from the positron annihilation, is usually measured as a whole.

In this paper, we propose a different detection method that could have the benefit of strongly reduced background, namely detecting IBD events using the tracking capabilities of a Time Projection Chamber (TPC) to fully reconstruct the ionization of each particle of the IBD final state. This is to be achieved using an organic liquid. It contains sufficient free protons to serve as a target for IBD reactions, as has already been proposed by previous studies \cite{McConkey2012,Dawson2014}. Specifically, in each IBD event one could seek to identify the initial positron track, the energy depositions of the two annihilation photons and of the delayed photon emitted due to the neutron absorption by a nucleus. Furthermore, ionization clusters induced by the neutron scattering off the nuclei of the detector material might be detectable.  

Reconstructing the tracks and showers from the original individual hits could provide information about the energy, the location and, with limitations, also the direction of the final state particles. 
The unequivocal signature of a completely resolved IBD final state could then enable background-minimized detection of electron antineutrinos. 
This would then allow to use the IBD antineutrino detection also for sources with smaller neutrino fluxes like nuclear waste or geoneutrinos and also in locations with less or even without any overburden. 

An additional benefit could be the access to information on the initial direction of the detected antineutrino. 
Liquid scintillation detectors have also looked for final state variables that contain information on the antineutrino direction \cite{Caden2012}. However, their position reconstruction usually lies in the order of few centimeters due to the time resolution achievable from the photodetectors surrounding the volume. Additional segmentation of the detector volume can further improve the position reconstruction and by that the directional sensitivity as discussed for SANDD \cite{SANDD}. 
Nonetheless, the results are based on the statistical analysis of a large number of antineutrino events. By using the final state as measured with a TPC, the variables sensitive to the antineutrino direction could be reconstructed with higher precision. New variables like the position of the neutron absorption or even neutron scattering with respect to the positron track become available and could be included in the reconstruction of the antineutrino direction.
Hence, the antineutrino direction could possibly be measured on an event-by-event basis.
If the directionality information can be improved sufficiently by this technique, one could be able to discriminate antineutrino background from the signal, e.g. reactor antineutrino background when measuring nuclear waste. The same holds for the determination of the distribution of radioactive nuclides within the rock formations when studying geoneutrinos.

In this paper, we present results of the first simulation-based statistical analysis of the expected IBD signatures in a liquid-organic TPC. The goal is to identify opportunities and challenges for their reconstruction. We use the example of antineutrinos emitted from nuclear waste as they present a significant challenge due to their energies reaching barely above the IBD detection threshold. Particular emphasis is placed on examining whether the antineutrino direction might indeed be reconstructable, as this would clearly be the most challenging reconstruction task. Compared to other experiments it is clear that there will be practical challenges to realize such detector. This study looks at potential capabilities of the general detection approach we present, in order to guide future experimental work to further assess the concept’s practical feasibility.

\section{Antineutrinos from Nuclear Waste}
\begin{figure}
    \centering
    \includegraphics[width=0.6\columnwidth]{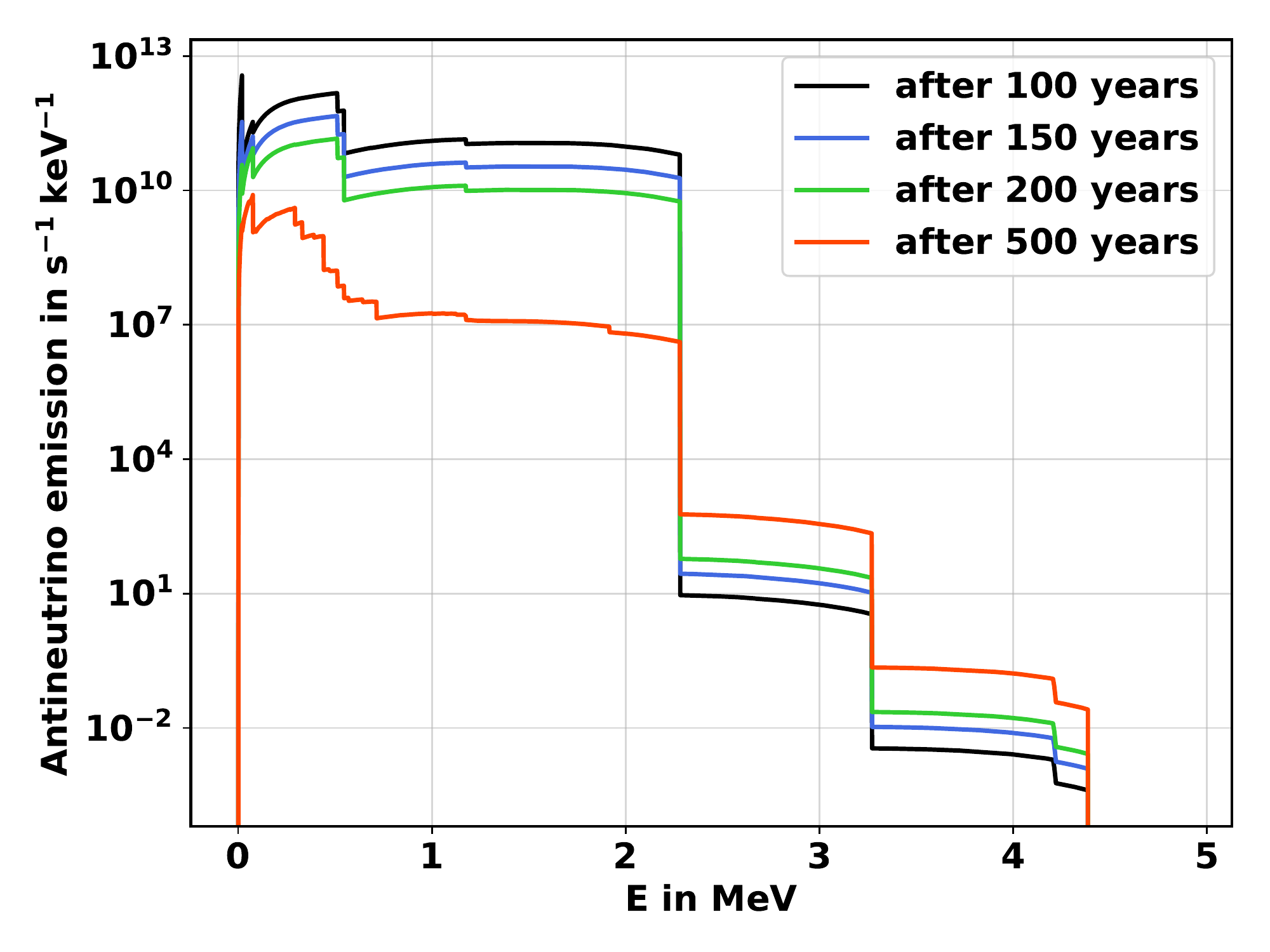}
    \caption{Energy spectra of antineutrinos emitted from one fuel assembly of a simulated PWR reactor with a burnup of $55\,\mathrm{MW d/kg}$ and enrichment of 4\%.}
    \label{fig:energySpectraWasteAntineutrinos}
\end{figure}

In the first few hundreds of years, the dominant radioactivity in nuclear waste comes from the $\upbeta$-decaying elements like ${}^{90}\mathrm{Sr}$ and ${}^{137}\mathrm{Cs}$. For the evaluation of antineutrino emissions, we obtained the isotopic concentrations of radioactive isotopes in spent fuel discharged from a pressurized water reactor, PWR, using a simulation with the reactor simulation code SERPENT2 \cite{Leppaenen2015}.
A burn-up of $55\,\mathrm{MWd/kg}$ and a uranium enrichment of 4\% was implemented. 
We then used the beta spectra of all beta-decaying nuclides from the software tool BetaShape \cite{betaShape} and converted them to the antineutrino spectra shown in figure \ref{fig:energySpectraWasteAntineutrinos} for different ages of the nuclear waste.

The antineutrinos are emitted at low energies with the emission significantly decreasing at $2.3\,\mathrm{MeV}$.
The maximum energy is reached at $4.3\,\mathrm{MeV}$ with an emission rate more than ten orders of magnitude smaller than for the lower energy range up to $2.3\,\mathrm{MeV}$.
Considering the detection via IBD with its threshold at $1.8\,\mathrm{MeV}$, the energy region that can be detected is fairly small. 
In this region, the antineutrino emission is dominated by the decay of the ${}^{90}\mathrm{Y}$ isotope, which is a decay product of the highly abundant fission product ${}^{90}\mathrm{Sr}$ and stays in equilibrium with it. 

Spent nuclear fuel is usually stored in a sealed containment that shields most of the ionizing radiation emitted from the waste. This shielding, however, has no impact on the antineutrino flux due to their low interaction cross sections. This makes the antineutrinos very interesting to be used to monitor the nuclear waste. Furthermore, the emitted antineutrino flux is directly depending on the waste mass, providing the opportunity for re-verification if the continuity of knowledge is lost. 

In \ref{sec:Appendix_ANueMonitoring} an exemplary scenario of a monitoring application for a geological repository is shown. It presents the expected event rates in an active medium of Tetra-Methyl-Silane as studied in this work. 

\section{Liquid-organic time projection chamber}
The TPC is a widely used tracking detector due to its good track reconstruction ability and energy loss resolution. 
It usually consists of a large detection volume in which charged particles deposit energy by ionizing the media and liberating electrons. Due to an externally applied electric field these ionization electrons drift towards an anode plane. 
A three-dimensional reconstruction is provided by a segmented, two-dimensional readout at the anode and the measurement of the drift time of the electrons.

Originally, TPCs use a gas mixture as sensitive material to measure the passage of charged particles. The neutral neutrino cannot be detected directly by an ionization detector, but charged final state particles from neutrino interactions with the detector material can be measured. Due to the extremely small cross section of neutrino interactions, a denser material than a gas is necessary.
For this reason, shortly after the invention of the gaseous TPC, the usage of liquid argon as active medium was suggested because of its higher density \cite{Rubbia1977}.
Nowadays, using liquefied noble gases has become an established technology for the detection of weakly interacting particles. 
In neutrino experiments like MicroBooNE~\cite{Acciarri2017} and ArgonNeuT~\cite{Anderson2012} liquid argon TPCs have been exploited, whereas liquid xenon is used in dark matter searches, e.g. the experiments XENON100~\cite{Aprile2012}, LUX~\cite{Akerib2013} and PandaX~\cite{Cao2014}. 
Currently, the DUNE neutrino experiment is building huge liquid argon TPCs for its far detector with a combined 70 kiloton of liquid argon as the target mass \cite{Acciarri2016}. First results with this technology were gained at the ProtoDUNE setup of the CERN neutrino platform~\cite{Abi2020}.

One drawback of liquid noble gases is that they need to be cooled down by a cryogenic system and therefore, it is very challenging to construct and operate a detector system with this technology. 
Hence, studying alternative high-density media that allow drifting electrons at room temperatures has been of special interest for a long time. One promising material is Tetra-Methyl-Silane, TMS, which is a non-polar organic liquid with spherically shaped molecules. 
Its usage in ionization detectors has been previously investigated \cite{Gonidec1988} with the focus on its application in calorimetry. 
For example, KASCADE-Grande~\cite{Engler1999} used liquid ionization chambers filled with TMS in their central calorimeter. 
In recent years, the use of these organic liquids as sensitive material of a TPC has been discussed \cite{McConkey2012,Dawson2014}. By operating small test chambers, the charge transport in organic liquids has been studied by measuring the charge yield \cite{Farradeche2018}, the drift velocity \cite{FAIDAS1989} and the mean free drift length \cite{Wu2020}. 
Still, these studies have to be intensified as open questions remain. For instance, further parameters relevant for the electron drift in organic liquids, such as the diffusion coefficients, still have to be investigated.

Another advantage of an organic liquid with respect to liquid noble gases is the presence of free protons due to its hydrogen content. 
Through this, a low threshold for the IBD process can be reached, making it possible to detect antineutrinos down to the energy of $1.8\,\mathrm{MeV}$.
Event rates should also be comparable with liquid scintillation detectors. 
For example, the number of target nuclei for IBD per volume in TMS is as large as 81~\% of the one in the scintillator material of KamLAND~\cite{KamLAND2008}.

As explained in the section above, the antineutrinos emitted from spent nuclear fuel have energies in the range between $1.8\,\mathrm{MeV}$ and $2.3\,\mathrm{MeV}$. Hence, liquid noble gases can not be used as target material to detect them via IBD due to their higher energy threshold of this reaction.

\section{Simulation setup and event generation}
In order to investigate the reconstruction of IBD events in a liquid-organic (LOr) TPC, we simulated antineutrino interactions within Tetra-Methyl-Silane as target material using the event generator GENIE~\cite{Andreopoulos2009} and the energy spectrum of figure\,\ref{fig:energySpectraWasteAntineutrinos} at 100 years.
Since the flux effectively vanishes above $2.3\,\mathrm{MeV}$, IBD can only take place at the hydrogen nuclei. The outgoing positron and neutron from the IBD reaction are then fed to a GEANT4~\cite{Agostinelli2003} simulation to investigate their interaction with TMS.
From this simulation we obtain the energy deposition of each particle. These are then translated into the number of electrons with an estimated electron yield $G_{if}$ that are used to simulate the spatial charge distribution at an anode with a readout structure of squared pads with a size of $5\,\mathrm{mm} \times 5\,\mathrm{mm}$ considering the drift velocity $v_{d}$ and the diffusion coefficient for longitudinal $d_{L}$ and transverse diffusion~$d_{T}$.

We examine two extreme, but still reasonable detector configurations with different maximum drift lengths of $l_{1} = 1\,\mathrm{m}$ and $l_{2} = 10\,\mathrm{m}$, respectively, where we keep a sensitive volume of $80\,\mathrm{m^3}$ congruous to our study presented in the \ref{sec:Appendix_ANueMonitoring}. For both configurations we assume a cathode high voltage of $-500\,\mathrm{kV}$ which generates a drift field of $E_{1} = 5\,\mathrm{kV/cm}$ and $E_{2} = 0.5\,\mathrm{kV/cm}$, respectively. For the first detector configuration, detector configuration~1, $G_{if,1} = 7\,\mathrm{e^{-}/keV}$ \cite{Holroyd2002} and $v_{d,1} = 5.5\,\mathrm{\upmu m/ns}$ \cite{Wu2020} are used. For the second configuration, detector configuration~2, we assume $G_{if,2} = 5\,\mathrm{e^{-}/keV}$ and $v_{d,2} = 0.5\,\mathrm{\upmu m/ns}$ using a mobility of $\mu = 105\,\mathrm{cm^2/V/s}$ \cite{Engler1996}.

Due to the lack of experimental data, we use the Einstein Relation ${D_{L,T}=\mu \cdot k_{B}T/\mathrm{e}}$ to estimate the thermal limit of the diffusion coefficient $D_{L,T}$ with given mobility $\mu$. To express the diffusion as a function of the drift distance $l_d$ instead of drift time $t_d$ via the formula $\sigma = \sqrt{2 \cdot D_{L,T} \cdot t_d} = d_{L,T} \sqrt{l_d}$ we transform the diffusion coefficient using equation \ref{eq:diffCoeff}:
\begin{equation}
    \label{eq:diffCoeff}
    d_{L,T} = \sqrt{\frac{2 D_{L,T}}{v_{d}}} = \sqrt{\frac{2k_{B}T}{\mathrm{e}E}}.
\end{equation}
Measurements of electron diffusion in liquid argon show that the actual diffusion coefficient $D_{L,T}$ is a factor 3--4 larger than the thermal limit \cite{Li2016} which can be explained by an energy increase of the electrons due to the external electric field \cite{shibamura1979}. Such an increase should also appear in TMS although measurements have to be performed to find out the magnitude. However, for simplicity we assume the same factor 4 for TMS and get a diffusion coefficient of  $d_{L,T, 1} \approx 60\,\mathrm{\upmu m/\sqrt{cm}}$ for detector configuration~1 and $d_{L,T, 2} \approx 190\,\mathrm{\upmu m/\sqrt{cm}}$ for detector configuration~2. We use the maximum drift length to study the maximum impact of diffusion which in the end limits to which extent tracks can be separated from each other. A summary of the simulation parameters can be found in table \ref{tab:simulationParameters}.

\begin{table}
    \centering
    \caption{Simulation parameters for the two detector configurations. In both cases a high voltage of $500\,\mathrm{kV}$ and a sensitive volume of $80\,\mathrm{m^3}$ is assumed.}
    \bigskip
    \begin{tabular}{l|c|c}
        
         & \parbox[c][40pt][c]{0.23\columnwidth}{\centering Detector \\ configuration 1} & \parbox[c][40pt][c]{0.21\columnwidth}{\centering Detector\\ configuration 2} \\
         \hline
        maximum drift length $l$ & $1\,\mathrm{m}$ & $10\,\mathrm{m}$ \\
        electric field $E$ & $5.0\,\mathrm{kV/cm}$ & $0.5\,\mathrm{kV/cm}$ \\
        drift velocity $v_d$ & $5.5\,\mathrm{\upmu m / ns}$ & $0.5\,\mathrm{\upmu m / ns}$ \\
        diffusion coefficient~$d_{L,T}$ & $60\,\mathrm{\upmu m/\sqrt{cm}}$ & $190\,\mathrm{\upmu m/\sqrt{cm}}$ \\
        electron yield $G_{if}$ & $7\,\mathrm{e^{-}/keV}$ & $5\,\mathrm{e^{-}/keV}$\\
        \hline
    \end{tabular}
    \label{tab:simulationParameters}
\end{table}

With these detector scenarios we want to evaluate two extreme, but still reasonable combinations of electric field and drift lengths.   
Detector configuration~1 has a higher electric field which comes with a higher drift velocity and smaller diffusion coefficient. Detector configuration~2 on the other hand has a larger drift length which results in less readout channels when keeping the same sensitive volume. Both scenarios are limited by the maximum high voltage. Even though we know that achieving this is challeging, we consider $500\,\mathrm{kV}$ as still manageable. In comparison, for the planned dual-phase detector of the DUNE project a cathode voltage of $600\,\mathrm{kV}$ was envisaged \cite{DUNE_TDR}. However, due to the design evolution of the ProtoDUNE dual-phase to Vertical Drift only $300\,\mathrm{kV}$ are needed \cite{ProtoDUNE_DP_yearlyProgress2021}. Their stability could be demonstrated in a HV campaign \cite{ProtoDUNE_DP_yearlyProgress2022}.

Another important point that has to be kept in mind is that in liquids, organics or cryogenic noble gases, high purity is needed to reach a sufficient free electron lifetime. For TMS this electron lifetime is constant for electric fields below $E = 20\,\mathrm{kV/cm}$ according to \cite{LOPES1988}. With the drift velocity rising proportional to the electric field a balance needs to be found between electric field and drift length. However, measurements have to be carried out to show if the required electron lifetimes are reachable. Electron lifetimes required for detector configuration 1 with the short drift length have already been shown \cite{OCHSENBEIN1988}. For detector configuration 2 with the long drift distance much larger electron lifetimes have to be achieved, comparable to the electron lifetimes in liquid Argon TPCs. Even though it is not yet proven if this can be achieved, the simulation of configuration 2 should show the influence of the diffusion. For this purpose these studies were performed with infinite electron lifetime. 

\section{Inverse beta decay signature in a LOr-TPC}
\begin{figure*}
    \centering
    \includegraphics[width=0.49\columnwidth]{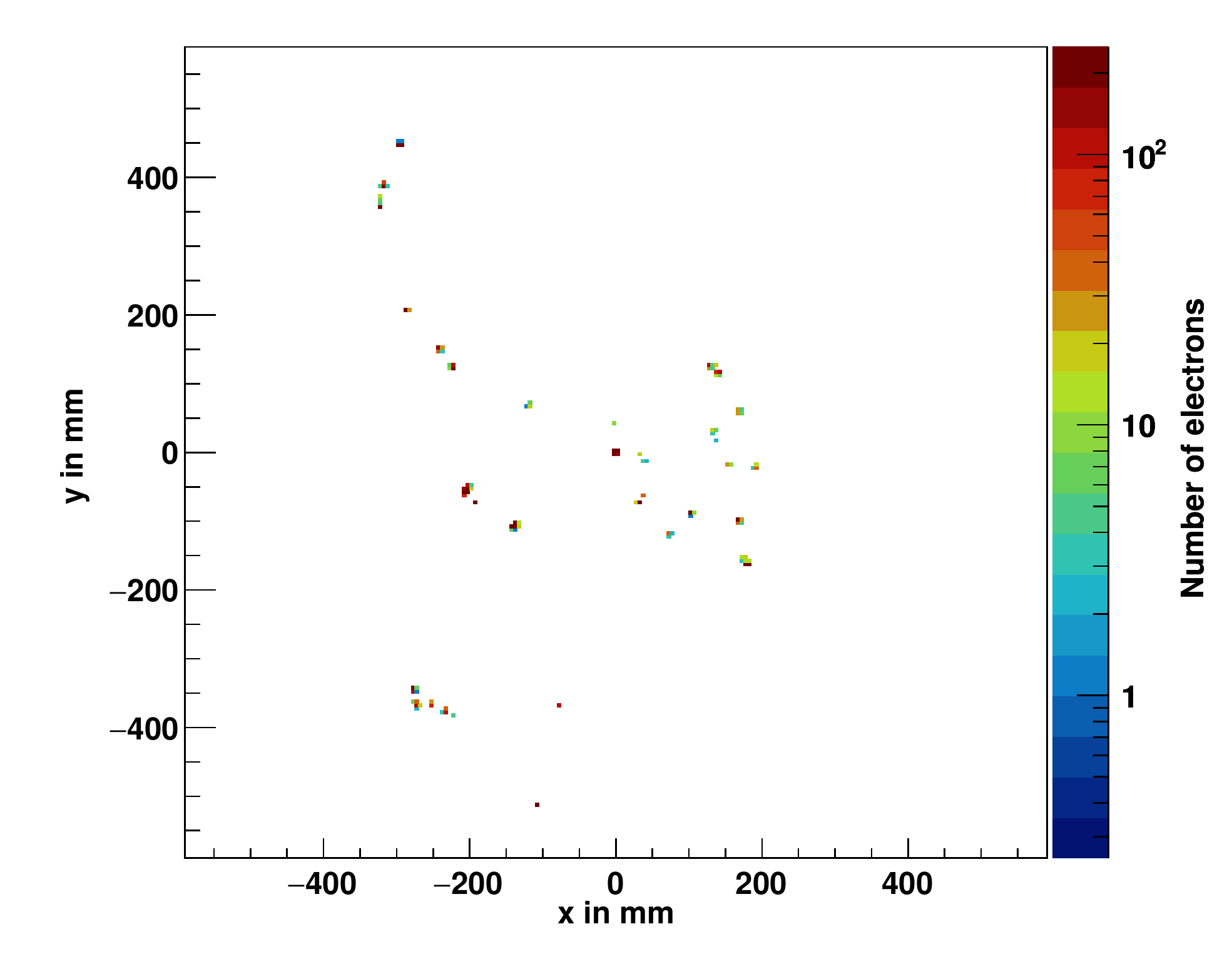}
    \includegraphics[width=0.49\columnwidth]{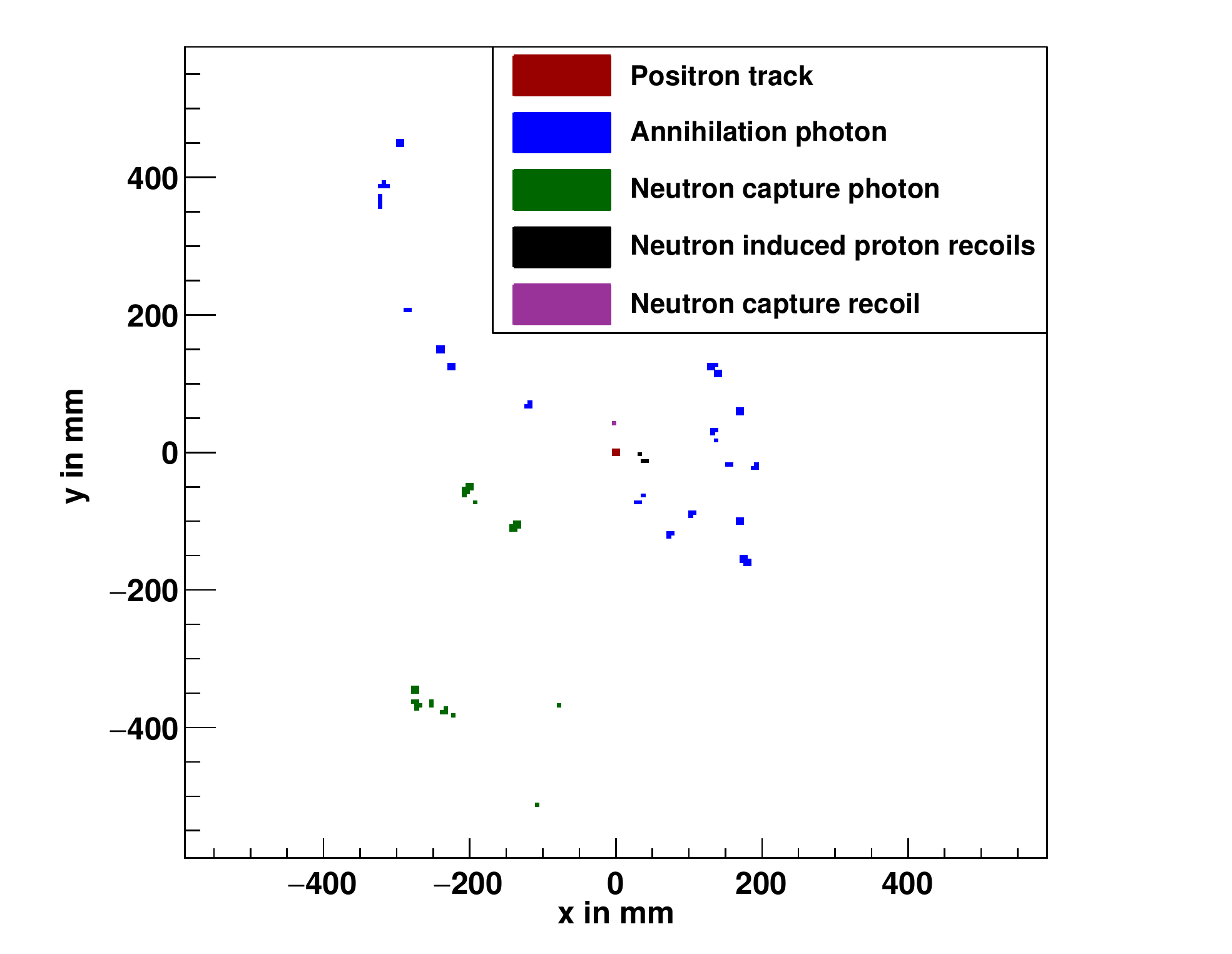}
    \caption{Simulation of an IBD event projected onto the anode for detector configuration~1. The left side shows the distribution of the single electrons in a color-coded scale. The right side shows which particle is associated to the ionization clusters.}
    \label{fig:eventDisplay}
\end{figure*}
Figure \ref{fig:eventDisplay} shows the interaction of an antineutrino entering the TPC perpendicular to the drift direction for detector configuration 1. It shows all energy depositions of the secondary particles projected on the anode integrated over the whole event time.
On the left side, for each readout pad the recorded ionization in number of electrons is shown. On the right side, the associated particle is mapped to each of the ionization clusters.
Contrary to liquid scintillation detectors, which can only distinguish a prompt from a delayed signal, here also the ionization of the positron and the scattering of the neutron is directly accessible. 

\begin{figure*}
    \centering
    \includegraphics[width=0.49\columnwidth]{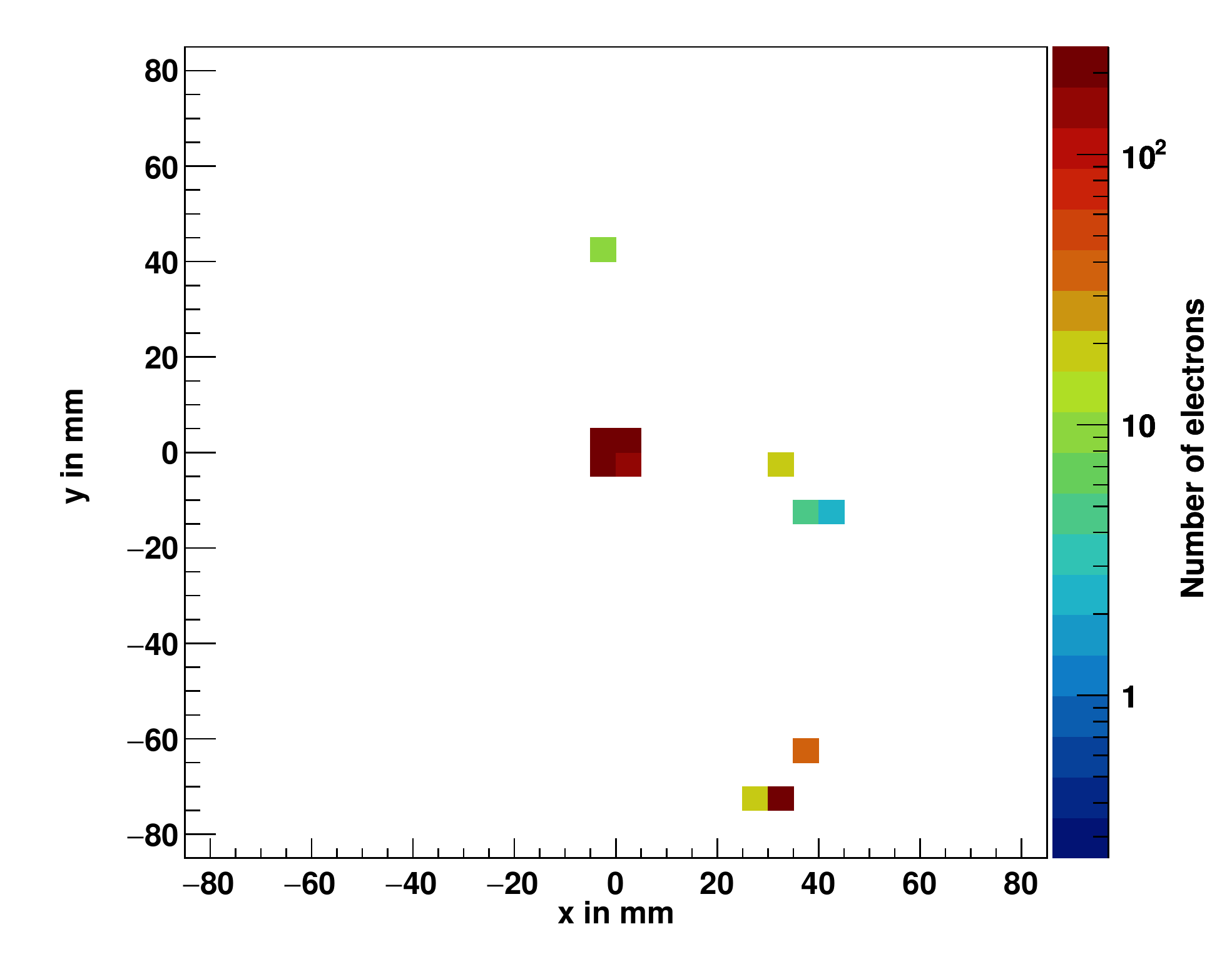}
    \includegraphics[width=0.49\columnwidth]{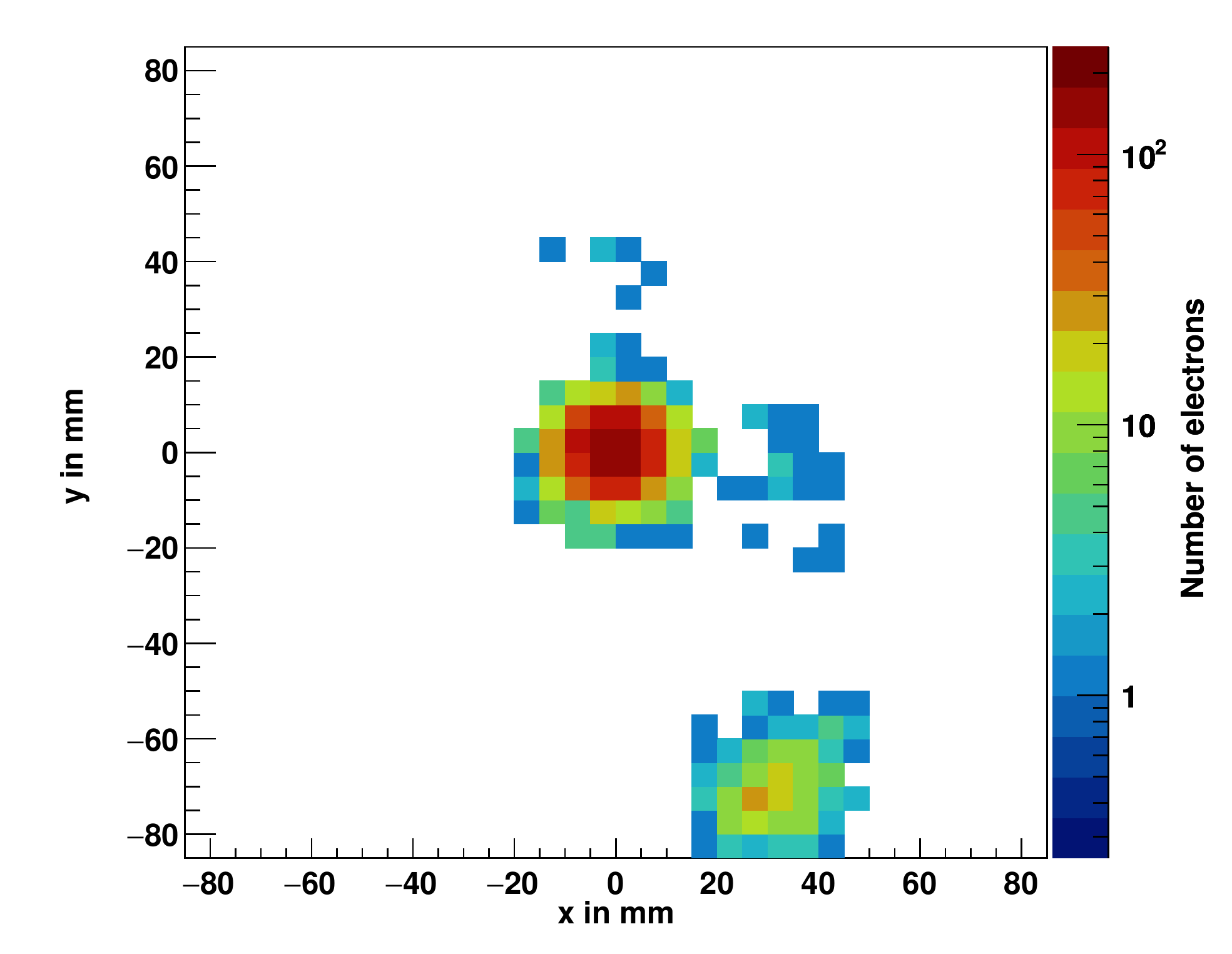}
    \caption{Comparison between the two detector configurations using the same event. The charge distribution of detector configuration~1 is shown on the left side. The right side shows the distribution of detector configuration~2. Both show the same zoomed-in view in the center of the event also shown in figure \ref{fig:eventDisplay}.}
    \label{fig:EventDisplay_Comparison}
\end{figure*}

For the comparison of the detector configurations, figure \ref{fig:EventDisplay_Comparison} depicts a zoomed view into the projected charge distributions simulated for the same event with the respective drift parameters. It shows the ionization cluster of the positron in the center, the ionization from the neutron-induced proton recoils on the right, the neutron capture recoil on top and some ionization of one of the annihilation photons on the bottom right. 
For detector configuration~2 that is shown on the right side of figure~\ref{fig:EventDisplay_Comparison}, the charge clusters are spread over more pads as expected due to the larger diffusion coefficient and the longer drift distance. As a result, the clusters of the positron and the neutron ionization are more likely to overlap and therefore, lead to a poorer separation. In addition, the larger spread of the charge reduces the potential electric pad signals that are especially low for the small ionization signals of the proton recoils which is slightly lower for detector configuration~2. 

\begin{figure}
    \centering
    \includegraphics[width=0.6\columnwidth]{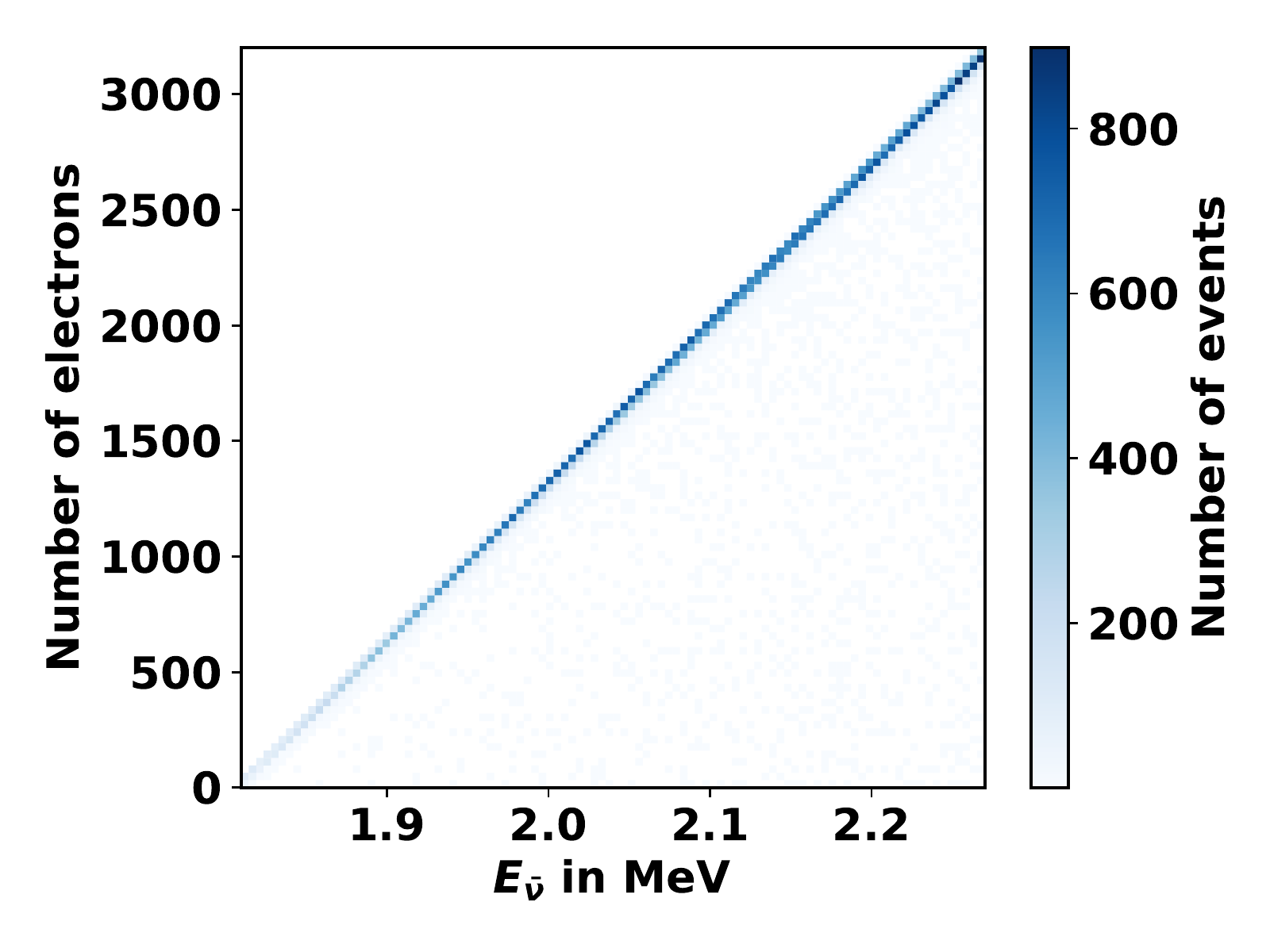}
    \caption{Number of ionization electrons created by the positron track in detector configuration~1 as function of the antineutrino energy. }
    \label{fig:posEnergyVsAntinueEnergy}
\end{figure}

We will explain each signal created by the particles of the IBD final state in the following. The numbers are given for detector configuration~1 first and for detector configuration~2 in brackets behind.
In the simulated energy region, the positron receives a kinetic energy smaller than $0.5\,\mathrm{MeV}$. This leads to a typical track length below $1\,\mathrm{mm}$ before annihilation. 
This track starts at the primary vertex of the IBD interaction and can be clearly identified by the up to $3200\,(2286)\,\mathrm{electrons}$ that are created along its path.
The ionization released along the positron track is a direct measure of the positrons' initial kinetic energy, which in turn measures the energy of the antineutrino as depicted in figure \ref{fig:posEnergyVsAntinueEnergy}.

At rest, the positron annihilates with an electron of the detector material and two $511\,\mathrm{keV}$ photons are created which can be used as a prompt signal to identify the IBD event as also done in liquid scintillation detectors. 
As shown in the event display, these photons produce characteristic energy depositions due to the fact that they initially travel a few centimeters losing their energy by Compton scattering before the final photoelectric effect takes place.

During its propagation, the neutron scatters off the nuclei of the detector liquid until it is thermalized and finally captured. 
The nucleon created by the neutron capture is produced in an excited state and finally releases the excitation energy via photon emission.
From 100,000 simulated events we found that in 93~\% of the cases the neutron is captured by a hydrogen nucleus which releases a single photon of $2.2\,\mathrm{MeV}$ energy. 
Again, this photon loses its energy by Compton scattering, traversing the liquid for a few centimeters. 
If the neutron is captured by another nucleus of the detector liquid,  several photons can be emitted with energies according to the excitation levels of the produced nucleus. 
On average it takes $(244.2 \pm 1.2)\,\mathrm{\upmu s}$ until the neutron is captured and the photon is released.
Detecting the photon from the neutron capture process in addition to the positron annihilation photons gives a clear event signature.

Even though the photons of the annihilation and the capture of the neutron are emitted at different times, the arrival time of the two charge signals at the anode can be ambiguous. First, the spatial distribution of the photon absorptions results in different arrival times of the ionization charge at the anode. Second, the electron cloud of this ionization diffuses while drifting to the anode. This can lead to an overlap in the arrival time of the signal making it harder to distinguish between the positron and the neutron signal. For detector configuration~1 28~\% of the events have electrons liberated by the neutron capture arriving at the readout before all electrons of the annihilation have arrived. For detector configuration~2 82~\% of the events show a concurrent arrival of the electrons due to the larger drift distance and the slower drift velocity. 
In most cases, the simultaneous arriving charge clusters, however, should be assignable to the different processes because of the spatial separation in the readout plane. Modern intelligent reconstruction algorithms could then reconstruct the events and may resolve the ambiguities, although this has to be studied in more detail.

The neutron produced by the IBD has typical kinetic energies between $1\,\mathrm{keV}$ and $3\,\mathrm{keV}$ due to its larger mass compared to the positron.
Therefore, the positron obtains practically all of the energy of the incoming antineutrino.
At the IBD energy threshold, the positron receives no kinetic energy and since the proton is at rest before the interaction, the neutron takes over the complete momentum of the antineutrino. Hence, it also has the identical direction as the incoming antineutrino. With increasing neutrino energy and thus rising positron momentum, the deviation between neutron and antineutrino direction increases due to momentum conservation \cite{Vogel1999}. Still, the neutron direction is largely correlated with the antineutrino direction at the considered energies. 
\begin{figure*}
    \centering
    \includegraphics[width=0.49\columnwidth]{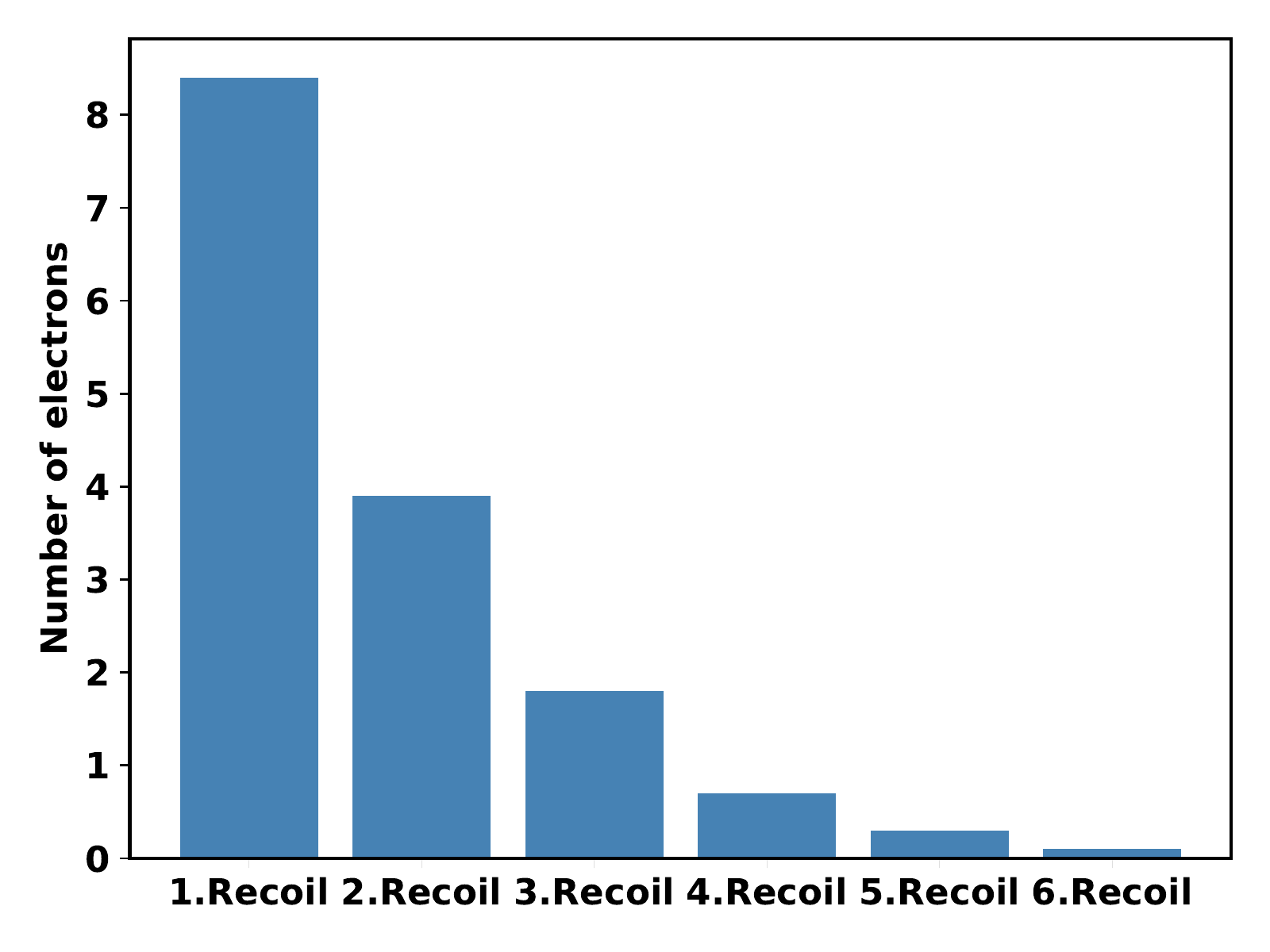}
    \includegraphics[width=0.49\columnwidth]{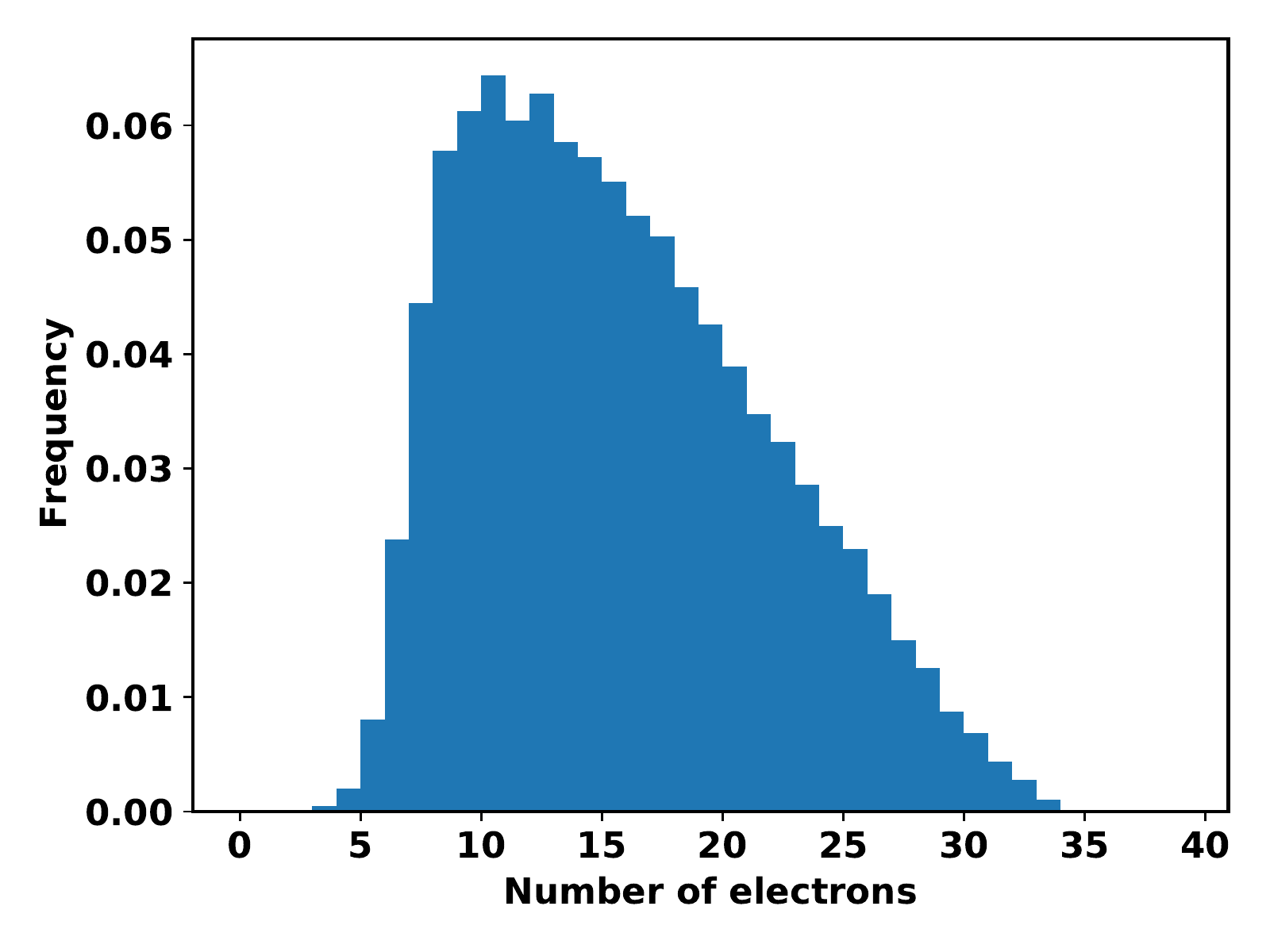}
    \caption{Number of electrons created by neutron-induced proton recoils in detector configuration~1. The left side shows the number of electrons created in each proton recoil. The right side shows the integrated number of electrons from proton recoils in each IBD event.}
    \label{fig:ProtonRecoilElectronNumber}
\end{figure*}

Afterwards, the neutron changes its direction due to scattering from the protons in the detection material. 
The mean distance between the antineutrino vertex and the first neutron-induced proton recoil is $(9.20 \pm 0.04)\,\mathrm{mm}$. 
Usually, the largest energy transfer occurs in the first collision. It decreases with each collision as shown on the left side of figure\,\ref{fig:ProtonRecoilElectronNumber}. 
The first proton recoil creates on average $8.4\,(6)$ electrons while it is only $3.9\,(2.8)$ for the second and $1.8\,(1.3)$ for the third. 
As the right side of figure\,\ref{fig:ProtonRecoilElectronNumber} shows, in each event an integrated signal of 5--30 (4--21) electrons can be expected from the neutron-induced proton recoils. The most probable value is around 10 (7). 

The various physical processes discussed so far yield very different amounts of ionization to be detected. Especially the detection of the neutron signal consists of only a few electrons and is therefore very challenging. However, different from dark matter experiments, this small signal is not be used as a trigger signal, but is intended to reconstruct a satellite charge cluster near the cluster created by the positron. To register the small charge amount of the neutron signal the concept of a dual-phase TPC can be utilized. Here, adding a gaseous layer on top of the liquid drift volume allows for electron amplification in the gas due to an avalanche process induced by a high electric field applied in this region. This concept of dual-phase TPCs has been successfully achieved in liquid xenon TPCs for dark matter searches \cite{Schumann2014}.

\section{Investigations on the reconstruction of the antineutrino direction}
An additional benefit of the full reconstruction of IBD final states could be the access to information on the initial direction of the detected antineutrino on an event-by-event basis. 

\begin{figure*}
    \centering
    \includegraphics[width=0.49\columnwidth]{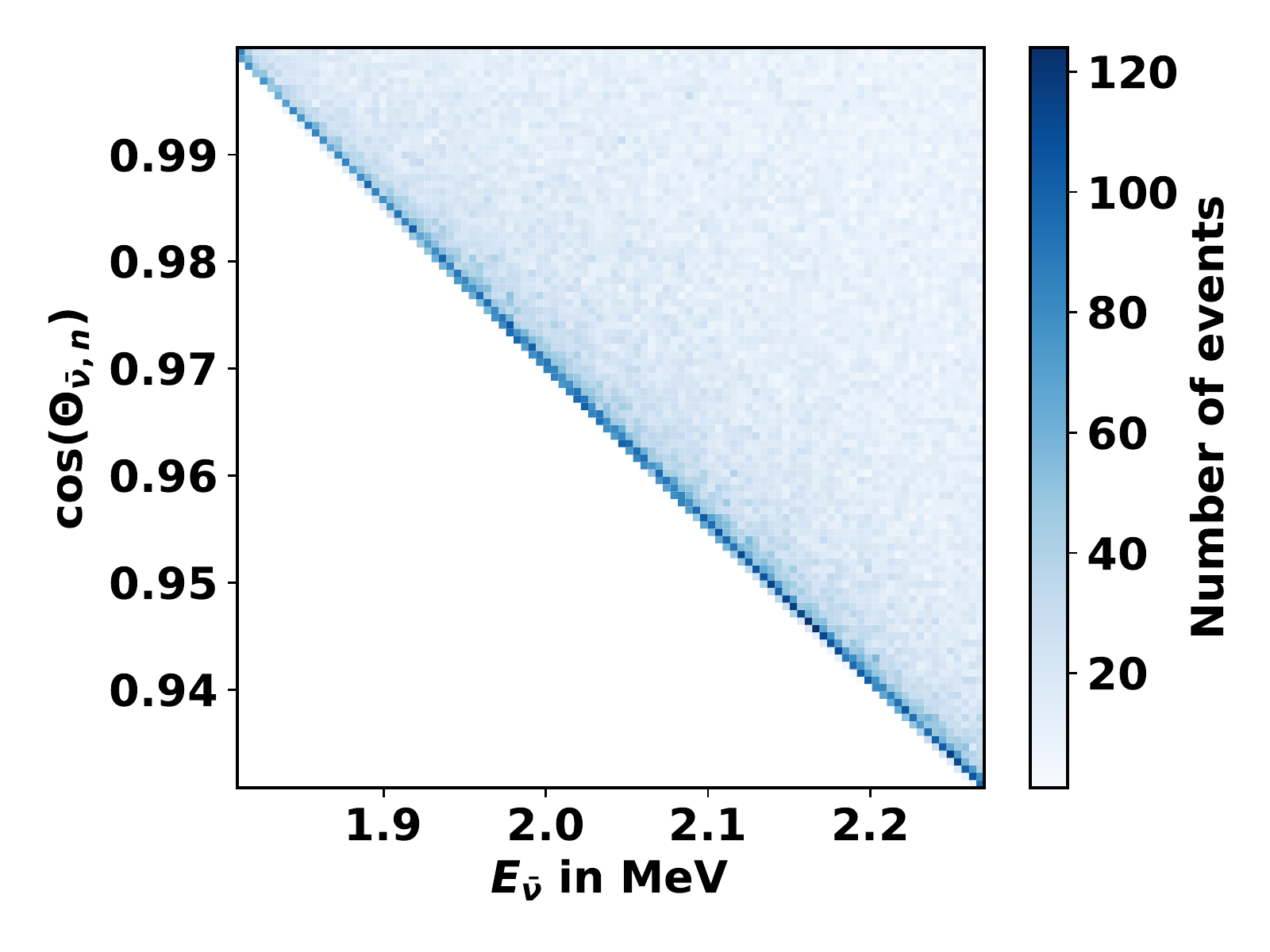}
    \includegraphics[width=0.49\columnwidth]{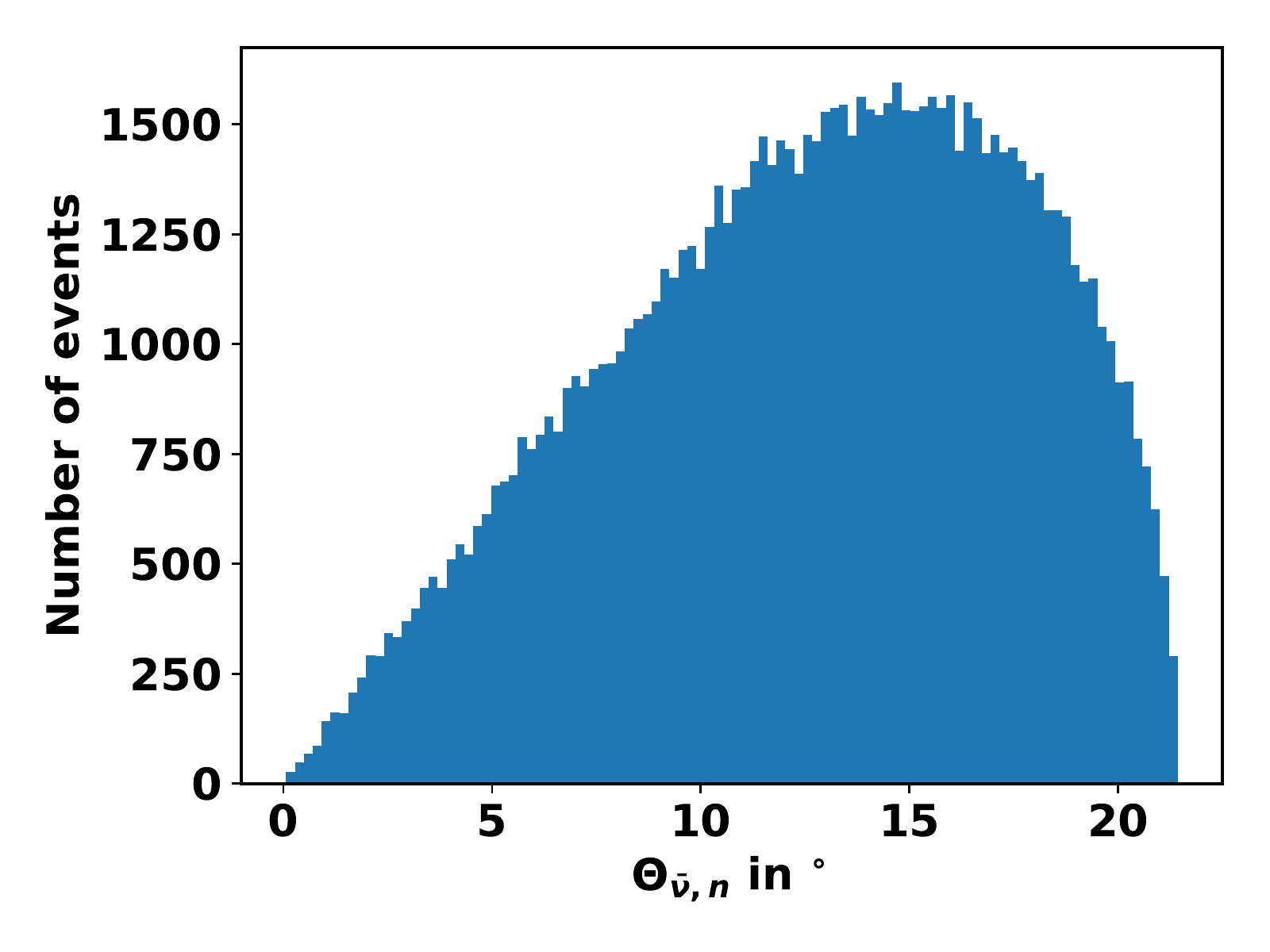}    
    \caption{Angular distribution of the neutron as a function of the incident antineutrino direction. The left side shows the angular deviation as a function of the antineutrino energy. The right side shows the angular probability distribution of IBD events based on the nuclear waste spectrum that we consider.}
    \label{fig:neutronAngleGenie}
\end{figure*}

We investigated IBD events simulated with GENIE for the nuclear waste antineutrino spectrum. The left side of figure \ref{fig:neutronAngleGenie} shows the angle between the incident antineutrino direction and the initial direction of the neutron as function of the antineutrino energy. Directly at the IBD threshold, the initial neutron direction is identical to the incident antineutrino direction. For higher antineutrino energies $E_{\nu}$ the deviation between the neutron and antineutrino direction increases.
The maximum angle can be described analytically, see equation \ref{eq:maxNeutronAngle} \cite{Vogel1999}.

\begin{equation}
    \label{eq:maxNeutronAngle}
    \cos{\phi_\texttt{max}} =\frac{\sqrt{2 E_{\nu} \Delta - (\Delta^{2} - m_{e}^{2})}}{E_{\nu}} \qquad \text{with}~\Delta = m_{n} - m_{p}.
\end{equation}

Since the most probable angle between neutron and antineutrino corresponds to the maximum angle (as can be seen in figure \ref{fig:neutronAngleGenie}), the neutron is initially directed in a cone around the antineutrino direction. Later the neutron undergoes several scattering events. The correlation with the initial neutron direction and the antineutrino direction is reduced by each scattering process. 
Therefore, it would be desirable to reconstruct the location of the first or of the first few neutron-proton scatterings. 
The connecting line between this location and that of the positron track would then yield a very good prediction of the direction of the antineutrino. 

On the right side of figure \ref{fig:neutronAngleGenie}, the angle between the neutron and antineutrino direction is shown for the IBD events based on the nuclear waste antineutrino spectrum considered by us. The maximum angle is $21^{\circ}$ and the distribution peaks at around $15^{\circ}$. 

\begin{figure}
    \centering
    \includegraphics[width=0.6\columnwidth]{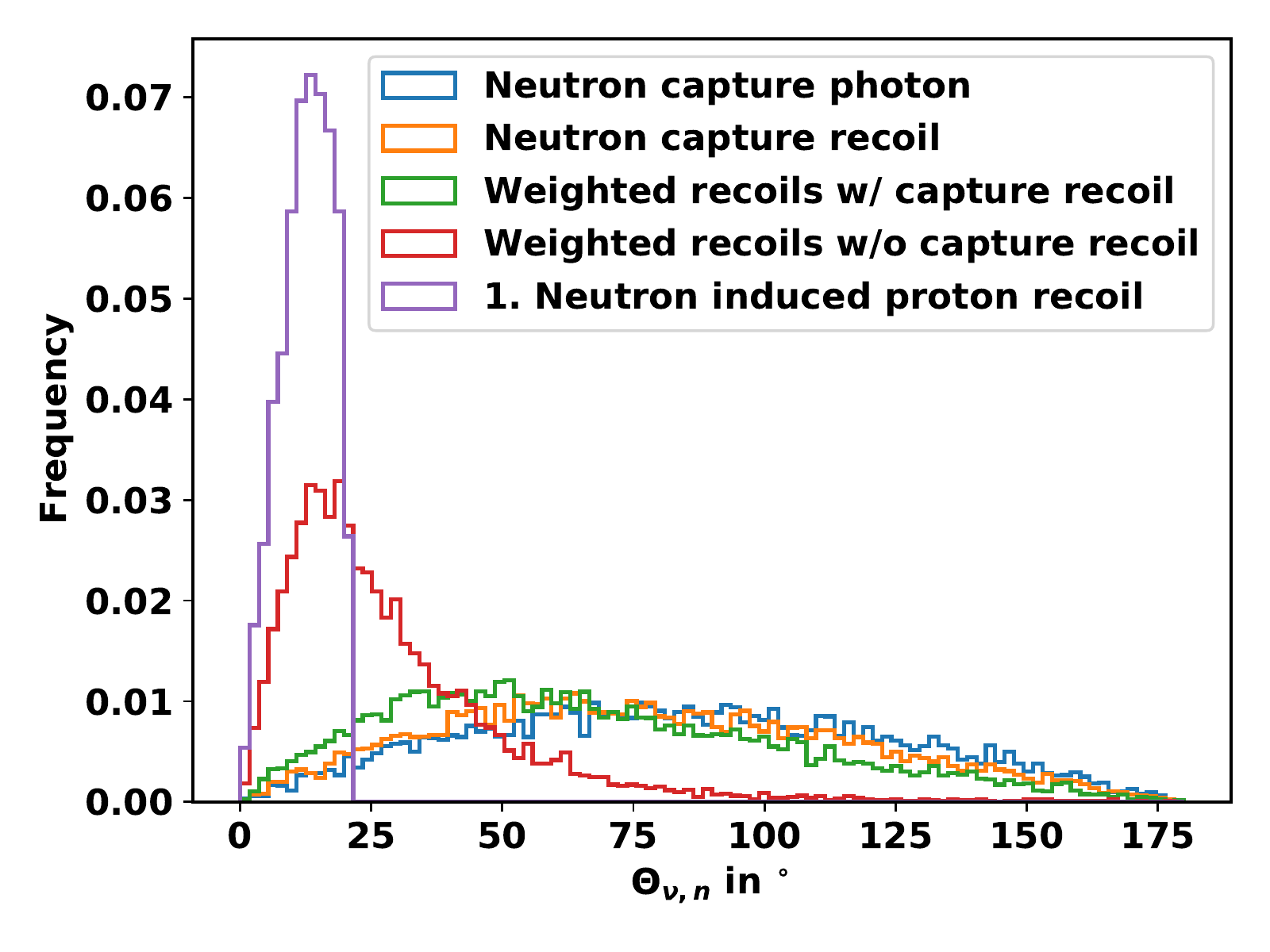}
    \caption{Angle between reconstructed neutron direction and incident antineutrino direction. Distributions are shown for different reconstruction methods, as explained in the text.}
    \label{fig:neutronReconstructionMethods}
\end{figure}

Furthermore, we studied the potential to reconstruct the antineutrino direction via the energy depositions using our GEANT4 Monte Carlo simulation. In this study, we considered five different definitions for a neutron direction vector that always starts at the IBD interaction origin and compared it with the incident antineutrino direction vector. 
We investigated (1) the first neutron-induced proton recoil, (2) the energy-weighted mean of all proton recoil ionization clusters, (3) all recoils including that of the nucleus capturing the neutron, (4) only the recoil of the nucleus capturing the neutron, and (5) the ionization of the first Compton-scatter of the photon released after the neutron capture. 
As depicted in figure \ref{fig:neutronReconstructionMethods}, the angular correlation between the direction of the initial antineutrino and the different definitions (1-5) diminishes significantly with more ionization energy taken into account. For (4) and (5) the distributions even become isotropic, so these events can not be used to reconstruct the direction on an event-by-event basis. 

Therefore, we conclude that for the best possible reconstruction of the antineutrino direction only the first neutron-induced proton recoil should be considered. However, as it might not be possible in all cases to isolate this energy cluster from the signature arriving at the anode, the second best option would be to use the energy-weighted position of all proton recoils without the recoil of the capturing nucleus.

As we have already shown above, the deposited energy on average decreases with each neutron scattering. Hence, automatically only the first few recoils will result in an measurable energy deposition and would be taken into account. 
As our simulations show, the capture recoil is usually $\Delta t \approx 240\,\mathrm{\upmu s}$ delayed and can therefore be excluded from the other proton recoils.

\begin{figure*}
    \centering
    \includegraphics[width=0.49\columnwidth]{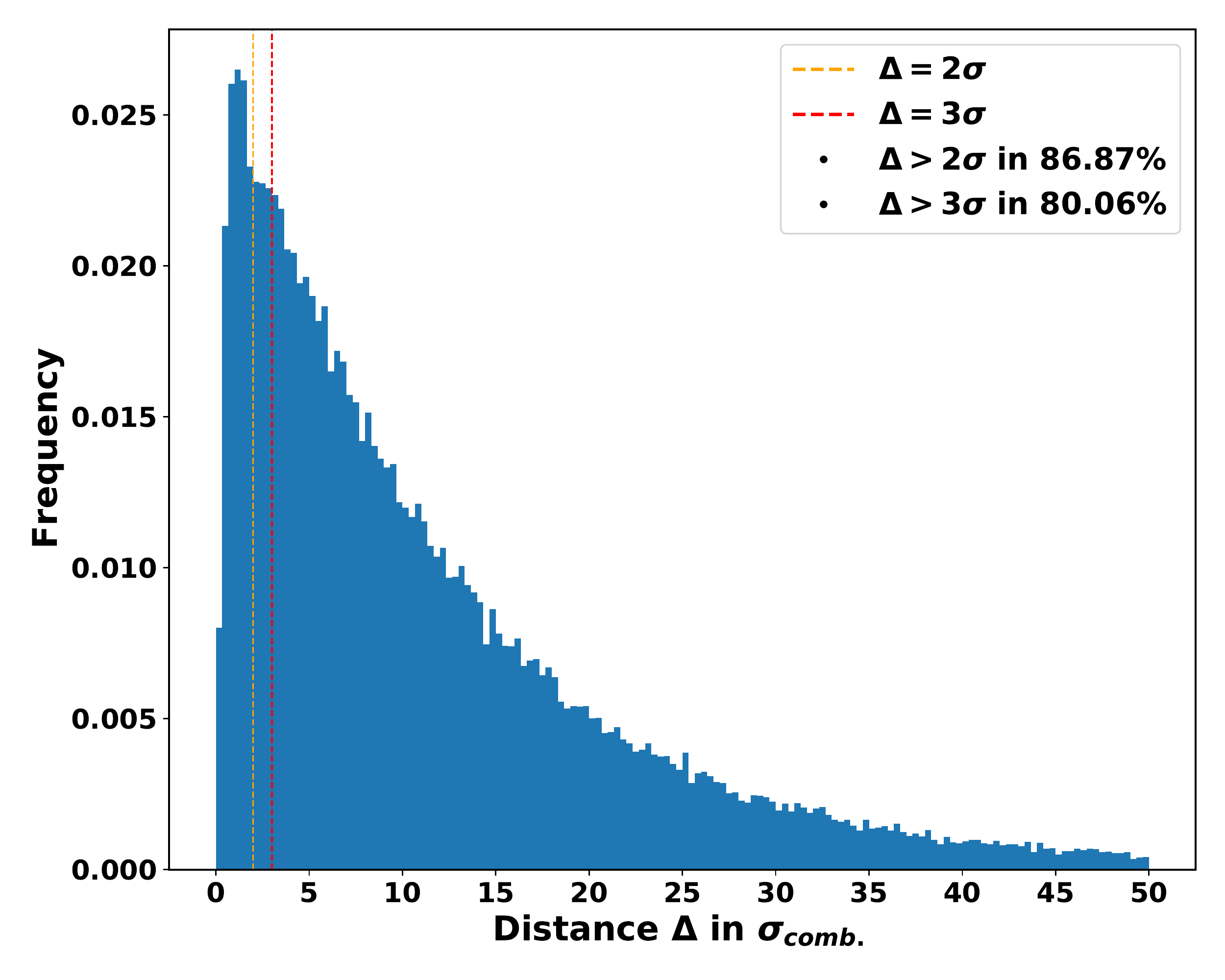}
    \includegraphics[width=0.49\columnwidth]{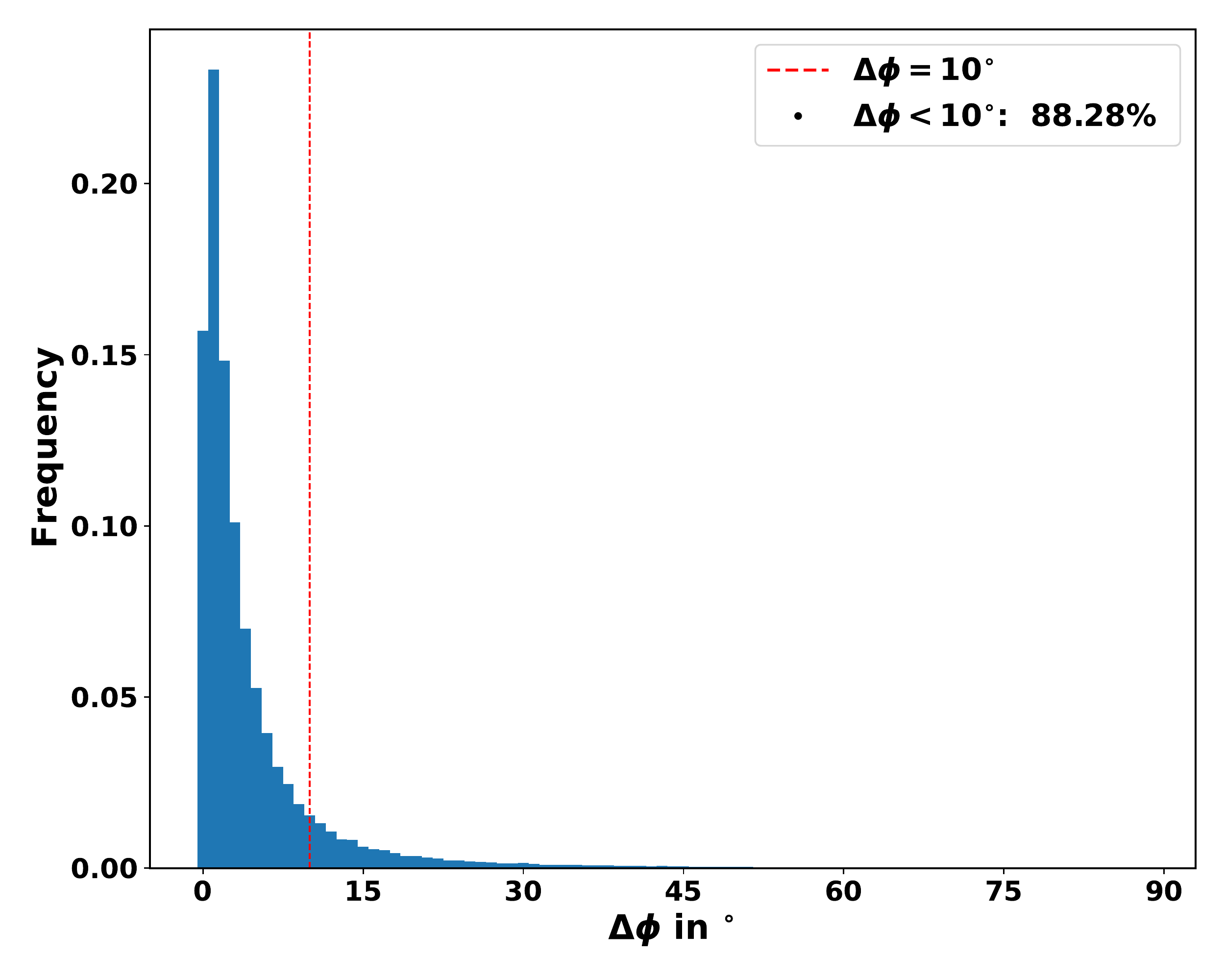}
    \caption{Distance of the charge distributions of neutron and positron on an anode for detector configuration~1 in combined standard deviations on the left side. On the right side the resulting angular deviation of the reconstructed and the true neutron direction for events with more than two combined standard deviation is shown. Here only the charge cluster of the first neutron-induced proton recoil was used.}
    \label{fig:NeutronPositronSeparation}
\end{figure*}

Finally, we studied the reconstruction of the neutron direction from the charge distribution created by the drifting electrons at an anode using the drift simulation described above. 
The question we pose is to which extent the neutron recoil cluster(s) can be separated spatially from the positron track, identifying them as small satellite clusters.  
Since the positron track is rather short, it can also be considered as a single ionization cluster from which we determine the center of charge and its two-dimensional standard deviation. 
We then determine the same for both the ionization clusters of all neutron-induced recoil protons and for only the first one. 
For IBD events where the antineutrino direction is perpendicular to the drift direction of the LOr-TPC we found that for 87~\% of the events, the charge distributions are more than two combined standard deviations apart in detector configuration~1 scenario as depicted in figure \ref{fig:NeutronPositronSeparation} using only the first neutron-induced proton recoil. 
For 88~\% of these events, the neutron direction can be reconstructed with a deviation of less than $10^{\circ}$ from the true neutron direction. 
When using all proton recoils for the neutron reconstruction, only 72~\% have a deviation of more than two combined standard deviation. In 68~\% of these events, the neutron direction can be reconstructed with less than a $10^{\circ}$ deviation.

For detector configuration~2, the charge distributions are only separated for 29~\% of the events with more than two standard deviations considering only the first neutron-induced proton recoil. From these events, 54~\% can be reconstructed with less than $10^{\circ}$ deviation. 
In case the charge distribution is reconstructed by all proton recoils 33~\% of the events are more than the two standard deviations apart in this scenario, of which  55~\% can be reconstructed with less than $10^{\circ}$ deviation.

Detector configuration~1 shows much better performance than detector configuration~2. This is mainly due to the larger diffusion in the case of detector configuration~2. On the one hand the diffusion coefficient is more than three times smaller than for detector configuration~2. On the other hand its maximum drift distance that we used for our evaluation is ten times larger. We additionally evaluated the separability for both detector configurations using a diffusion coefficient of $d_{L,T} = 300\,\mathrm{\mu m/ \sqrt{cm}}$. For detector configuration~1, this reduces the number of events that can be separated with more than two combined standard deviations to 
49~\% which would still be in an acceptable range. Due to the large drift distance in detector configuration~2, the clusters smear out that much, that they can not be separated anymore. We conclude that measurements of the diffusion coefficients are needed to specify a maximum drift length. 

For our example, assuming a detector volume of dimensions $10\times4\times2\,\mathrm{m^3}$, the setup of detector configuration~1 with the $1~\mathrm{m}$ maximum drift length can be realized by a central cathode in parallel to the $10\times 4\,\mathrm{m^2}$ large side. This setup would require a total readout area of $80\,\mathrm{m^2}$. The setup of detector configuration~2 could be realized with a cathode on one of the $2\times4\,\mathrm{m^2}$ large sides. In summary, detector configuration~1 delivers better results due to the ten times shorter drift length, but needs a ten times larger readout area instrumented. 

\section{Conclusion and Outlook}
We presented the first simulation-based statistical analysis of the expected IBD signatures in a liquid-organic TPC. This is a newly proposed detection method for neutrino scattering events using a liquid-organic TPC which aims for a full reconstruction of the IBD final state.
The identification of this unequivocal event signature has the potential to enable a background-minimized detection of antineutrino events.
The basis for this is the reconstruction of the positron track. On the one hand, it improves the neutrino energy resolution due to the correlation of ionization of the positron track to the initial antineutrino energy. On the other hand, it makes it possible to distinguish the IBD events containing a positron signal plus two annihilation photons from background events involving $\upbeta$-electrons.

Based on our results, we conclude that it could be possible to indeed associate the energy depositions to the final state particles and achieve a complete reconstruction of IBD events. Further studies on the identification with dedicated reconstruction algorithms have to be performed. Recent advances in these analysis techniques are promising in this respect. 

In addition, we showed that the direction of the incoming antineutrino could be estimated from the locations of the energy depositions caused by the recoil protons produced by the scattering of the final state neutrons. Even though their ionization signal is very small, it might be detectable as a satellite cluster to the positron track. Experimental studies have to be undertaken to decide whether this is achievable, given the fact that the positron track's signal is two orders of magnitude larger than that of the recoil protons. Additionally, the separability is strongly depending on the diffusion coefficients and the drift distance. Since no measurement data for the diffusion coefficient are available, we could only estimate them for our studies. The concept could lead to a measurement of the antineutrino direction on an event-by-event basis, even when our simulations show that not all proton recoils have a sufficient distance from the positron track. It would in any case, however, be a large improvement compared to current capabilities using scintillation detectors.

Future simulation studies should also address signatures for antineutrinos entering the detector in parallel to the electron drift direction. Here, the neutrons and positron tracks would be much closer together in the anode plane. How they could be distinguished in a three dimensional reconstruction also using the drift time measurement should be further assessed. 

Finally, before a liquid-organic TPC as it is considered in our studies can be realised, some technical specifications still have to be demonstrated experimentally. Most important is to reach electron lifetimes that allow drift lengths in the order of meters. To achieve this goal, the liquid needs to be purified to a ppb-level of oxygen. Here, methods deployed at the purification of liquefied noble gas TPC can be utilized. 
Furthermore, to realise a dual-phase readout for the signal registration, the extraction of electrons from the liquid into a gas needs to be shown. Otherwise, only the larger signals from the IBD signature can be detected.

We will soon start further studies based on a TPC prototype to investigate the technical realisation of a detector described in the simulation studies in this paper. We intend to measure drift velocity, electron lifetime and the diffusion parameters of TMS. Additionally, we want to examine the extraction of electrons from the liquid into a gaseous amplification structure.

\section*{Declaration of Competing Interest}
The authors declare that they have no known competing financial interests or personal relationships that could have appeared to influence the work reported in this paper.

\section*{Acknowledgments}
We thank Madalina Wittel for her engagement in starting this project and we acknowledge the contributions of David Betram, Helge Haveresch, Anike Ohm and Hagen Weigel. Additionally, we thank Yan-Jie Schnellbach for fruitful discussions about this publication. The project was funded by the Federal Ministry of Education and Research (BMBF) and the Ministry of Culture and Science of the German State of North-Rhine Westphalia (MKW) under the Excellence Strategy of the Federal Government and the Länder [funding code OPSF 573]. Further support came from the Federal Ministry for Economic Affairs and Energy [funding code 02W6281]. Malte Göttsche and Thomas Radermacher are generously funded by the VolkswagenStiftung Freigeist Fellowship.

\appendix
\section{Initial study for monitoring a geological repository with a liquid-organic TPC}
\label{sec:Appendix_ANueMonitoring}
As an initial study we investigated a model loosely related to the now discarded Gorleben site \cite{DBETec12} in Germany. It comprises 2240 containers with ten spent fuel assemblies each. The production year of the fuel was simulated corresponding to the German program, though we assumed the last fuel to be discharged in 2020. 

The layout foresaw the storage caverns to be located at a depth of $870\,\mathrm{m}$. In our assessment, we envisaged that an antineutrino detector would be placed at the same depth as the repository, but $50\,\mathrm{m}$ away from the outermost containers. In view of the transportation and deployment requirements, we considered a detection volume of $80\,\mathrm{m^3}$ of Tetra-Methyl-Silane for the antineutrino detector. The results of this preliminary study are summarised in table \ref{tab:appendix_GorlebenExample}. It can be seen that the expected detection rate diminishes significantly with time due to the decay of ${}^{90}\mathrm{Sr}$. Note that larger detectors or multiple detectors at different locations could be employed, including above the repository \cite{schnellbach2023}. These results show that antineutrino monitoring with a liquid-organic TPC could be possible, but indeed only using a detection technique with high background suppression. 
\begin{table}
    \centering
    \caption{Expected detection rates of the antineutrino emissions without background evaluated with a detector volume of $80\,\mathrm{m^3}$ TMS in a TPC.}
    \bigskip
    \begin{tabular}{|c|c|}
        \hline
       \parbox[c][40pt][c]{0.25\columnwidth}{\centering Time elapsed since 2020 in years} & \parbox[c][40pt][c]{0.25\columnwidth}{Number of IBD events detected in 3 months} \\
        \hline
        10 & 761 \\
        25 & 532 \\
        50 & 293 \\
        100 & 89 \\
        125 & 49 \\
        150 & 27 \\
        \hline
    \end{tabular}
    \label{tab:appendix_GorlebenExample}
\end{table}

\bibliographystyle{elsarticle-num}
\bibliography{bibliography}

\end{document}